\documentstyle[12pt]{article}
\input epsf.tex

\makeatletter
\def\vereq#1#2{\lower3pt\vbox{\baselineskip1.5pt \lineskip1.5pt
\ialign{$\m@th#1\hfill##\hfil$\crcr#2\crcr\sim\crcr}}}
\makeatother

\begin{document}
\setlength{\unitlength}{1mm}
\textwidth 15.0 true cm
\textheight 22.0 true cm
\headheight 0 cm
\headsep 0 cm
\topmargin 0.4 true in
\oddsidemargin 0.25 true in

\def\beq{\begin{equation}}   \def\eeq{\end{equation}}

\newcommand{\gsim}{\lower.7ex\hbox{$\;\stackrel{\textstyle>}{\sim}\;$}
}
\newcommand{\lsim}{\lower.7ex\hbox{$\;\stackrel{\textstyle<}{\sim}\;$}
}

\newcommand{\ra}{\rightarrow}
\newcommand{\ve}[1]{\vec{\bf #1}}

\newcommand{\La}{\overline{\Lambda}}
\newcommand{\Lam}{\Lambda_{QCD}}
\newcommand{\re}[1]{Ref.~\cite{#1}}

\newcommand{\eq}[1]{Eq.\hspace*{.1em}(\ref{#1})}
\newcommand{\eqs}[1]{Eqs.\hspace*{.1em}(\ref{#1})}
\newcommand{\Eq}[1]{Eq.\hspace*{.1em}(\ref{#1})}
\newcommand{\Eqs}[1]{Eqs.\hspace*{.1em}(\ref{#1})}
\newcommand{\PSbox}[3]{\mbox{\rule{0in}{#3}\includegraphics{#1}\hspace{#2}}}

\renewcommand{\Im}{\mbox {Im}\:}

\newcommand{\be}{\beta}
\newcommand{\ga}{\gamma}
\newcommand{\de}{\delta}
\newcommand{\al}{\alpha}
\newcommand{\as}{\alpha_s}

\newcommand{\GeV}{\,\mbox{GeV}}
\newcommand{\MeV}{\,\mbox{MeV}}

\def\lsim{\mathrel{\rlap{\lower3pt\hbox{\hskip0pt$\sim$}}
    \raise1pt\hbox{$<$}}}         %less than or approx. symbol
\def\gsim{\mathrel{\rlap{\lower4pt\hbox{\hskip1pt$\sim$}}
    \raise1pt\hbox{$>$}}}         %greater than or approx. symbol

\begin{titlepage}
\begin{center}
~{} \hfill    CERN-TH-99-219\\
~{} \hfill SU-ITP-99/38\\
~{} \hfill LBNL-44031\\

\vskip .15in

{\large \bf Logarithmic Unification From Symmetries\\
Enhanced in the Sub-Millimeter Infrared}
\footnote{Supported in part by the DOE under grant 
DE-AC03-76SF00098, the NSF under grant PHY-95-14797
and by the Alfred P. Sloan Foundation.  Email: {\tt nima@thwk1.lbl.gov},
{\tt savas@leland.stanford.edu}, {\tt j@marchrussell.org}}

\vskip 0.25in

{\bf Nima Arkani--Hamed$^{**}$, Savas Dimopoulos$^{+}$,\\ and John
March-Russell$^{\dagger}$}

\vskip 0.2in

$^{**}$ Department of Physics, University of California, Berkeley, CA 94530\\
Theory Group, Lawrence Berkeley National Lab,
Berkeley, CA 94530
\vskip 0.15in
$^{+}$ Physics Department, Stanford University, Stanford, CA 94305
\vskip 0.15in
$^{\dagger}$ Theory Division, CERN, CH-1211, Geneva 23, Switzerland

\end{center}

\begin{abstract}In theories with TeV string scale and sub-millimeter
extra dimensions the attractive picture of logarithmic
gauge coupling unification at $10^{16}$ GeV is seemingly
destroyed.  In this paper we argue to the contrary that 
logarithmic unification {\it can} occur in such theories.
The rationale for unification is no longer that a
gauge symmetry is restored at short distances, but rather that a geometric
symmetry is restored at large distances in the bulk away from our 3-brane.  The
apparent `running' of the gauge couplings to energies far above the
string scale actually arises from the logarithmic variation of classical fields
in (sets of) two large transverse dimensions.  We present a number of $N=2$ and
$N=1$ supersymmetric D-brane constructions illustrating this picture for 
unification.
\end{abstract}

\end{titlepage}

\newpage

\section{Replacing the Desert with the Bulk}

It has recently been realized that the fundamental scales of
gravitational and string physics can be far beneath the
conventional energies of $10^{17}-10 ^{18}$ GeV. This can be
accomplished if $n$ spatial dimensions of size $\sim R$ are much
larger than the fundamental Planck/string scales $\sim M_s$. At
distances much larger than $R$, the weakness of gravity can be
understood from Gauss' law, which relates the effective $4d$
Planck scale to the fundamental scale via
\begin{equation}
M_{pl}^2 \sim R^n  M_{s}^{2 + n}
\end{equation}
The original motivation was to bring the fundamental scales close
to the weak scale, in order to solve the hierarchy
problem~\cite{ADD,AADD,ADDlong,AHDMR}. Putting $M_s \sim$
TeV, the radius $R$ ranges from $\sim$ mm for $n=2$ to $\sim (10
{\rm MeV})^{-1}$ for $n=6$. These large dimensions are not in
conflict with experiment if the SM interactions are confined to a
3-brane in the extra dimensions.

Perhaps the most surprising aspect of this framework is that,
despite its profound modifications of physics at both
sub-millimeter and TeV scales, it is experimentally viable,
non-trivially surviving laboratory, astrophysical and cosmological
constraints~\cite{ADDlong}.  The last year has also seen the
growing realization that the large space in the extra dimensions
replaces the old ultraviolet desert as the new arena in which to
address other outstanding mysteries of the Standard Model, such as
the origin of flavor and absence of FCNC's \cite{AHD},
small neutrino masses \cite{ADDMR},
and proton stability~\cite{ADDlong,AHD,AHS}. The dynamics of
the bulk also plays an important role in the early universe
cosmology of this framework\cite{cosmo}.

One important and general lesson has been that there are new,
intrinsically higher-dimensional mechanisms for generating small
parameters in the four dimensional theory, the smallness of which
are not guaranteed by symmetries of the low energy theory.
Instead, higher dimensional locality can guarantee that
interactions between fields separated in the extra dimensions are
suppressed. In this way, if symmetries are broken at $O(1)$ on
distant branes, with the breaking transmitted to our brane by
massive bulk fields, the breaking can be exponentially small on
our wall \cite{AHD}. In another context, this effect helps alleviate 
the SUSY flavor problem in models of wall to wall SUSY breaking \cite{RS1}.
As another example, proton decay can be suppressed if
quarks and leptons are slightly `split' in the extra
dimensions~\cite{AHS}. 

In order to go further in making this framework as compelling as
the standard picture with the string scale $\sim 10^{18}$ GeV
and low-energy supersymmetry (SUSY) breaking, two important
theoretical issues must be addressed:
\begin{itemize}
\item
Why are the radii of the extra dimensions so large
compared to the string scale?
\item
What about gauge coupling unification?
\end{itemize}

In this paper, we will focus on the issue of gauge coupling
unification~\cite{DDG}, and argue that the picture of logarithmic gauge
coupling unification\footnote{Approaches to gauge coupling unification
with power-law running have been discussed in ~\cite{unif}.}
may still emerge in string theory with a low
string scale and large extra dimensions. As we will see, 
this provides another illustration
of the way in which the bulk can mimic physics associated with the old Desert.
Most of the formal results we discuss below are well known in the literature
on D-brane constructions of gauge
theories~\cite{Dbrane,HanW,MQCD,LPT,DL,matt,orb,karch}, although
we will be interested in going beyond the `decoupling' limit.
What we wish to point out is the possible application of these
results to phenomenology. 

There have recently been a number of papers 
addressing the possibility of `running' far above the string scale 
\cite{B,AB,earlier}. Running the couplings above the string scale is however
not enough, we need a reason for the couplings to {\it unify} far above the
string scale. This is challenging because in string theory, the couplings 
usually unify at the string scale. We will argue that the new rationale for 
unification far above the string scale can be an enhanced geometrical symmetry
at large distances away from our 3-brane in the bulk. We also further clarify
the conditions under which the couplings can `run' above the string
scale in D-brane configurations with $N=2$ and $N=1$ supersymmetry. 

\section{Logarithmic `Running' from the Infrared}

There are two remarkable features of the usual picture of gauge
coupling unification in supersymmetric extensions of the Standard
Model: first that the couplings unify at all, and second that they
unify so close to the String/Planck scale. Of course, given the
precision with which the couplings have been measured, the near
miss of the unification and string scales is usually considered a
problem, but at zeroth order it is remarkable that the naive
scales of gauge and gravitational unification are so close to each
other. For the moment, we will ignore the difference between the
GUT and Planck scales, and will return to this point
later.

Of course, we have not actually {\it measured} the gauge couplings
at energies approaching the GUT scale; all we know is that the
measured strength of the gauge couplings $\alpha_i^{-1}$ at low
energy satisfy, to high accuracy, the relation

\begin{equation}
\alpha_i^{-1}(\mbox{TeV}) = \alpha_0^{-1} - \frac{b_i}{2 \pi}
\mbox{log}\left(\frac{M_{pl}}{\mbox{TeV}}\right). \label{unif}
\end{equation}

Can this relationship possibly be reproduced in a theory
with a low string
scale $M_s\simeq {\rm TeV}$ and large extra dimensions? In
particular, we wish to reproduce the picture of {\it logarithmic}
gauge coupling unification, as opposed to power-law unification
\cite{DDG}. Even if power-law running explains why the couplings
unify, it can not explain why the unification appears to happen so
close to the Planck scale. Put another way, suppose that instead
of unifying at $\sim 10^{16}$ GeV, the couplings were found to
unify at $10^8$ GeV; power-law running could still accommodate
this. The link between the unification scale and the gravitational
scale in Eqn.(\ref{unif}) would then be wholly accidental. We do
not wish to view this link as an accident, and will therefore try
to reproduce Eqn.(\ref{unif}) in the context of theories with
large extra dimensions.

Of course naively, once we hit the (low) string scale $M_s$, we
can't continue to `run' past the TeV scale using the RGE's of the
4-dimensional effective field theory, and there is no source for a
UV logarithm of magnitude  $\log(M_{pl}/M_s)$.  In particular,
scattering experiments performed at $\sqrt{s} \gg M_s$ no longer
see just a gauge theory on our brane decoupled from the full set
of string modes and from the bulk degrees of freedom, so there is
no sense in speaking of usual QFT `running' at these scales.  However,
the usual $M_{pl}$ does exist
as a physical scale; it is set by the size of
the bulk. In units where $M_s = 1$, $M_{pl} \sim \sqrt{V_{bulk}}$.
So, if the gauge coupling on our brane can have logarithmic
sensitivity to the volume of the bulk, there is a hope that the
correct magnitude logarithm is available. Notice that the source
of the logarithm in this picture is an {\it infrared} effect.

Under what conditions can quantities on the brane depend on the
size of the bulk \cite{AB}? Clearly, the brane must couple to light fields
that can propagate in the bulk in order to have any sensitivity to
the volume of the extra dimensions. Gravity is one
model-independent field that must propagate in all $n$ large extra
dimensions. There may be other light fields as well, which can
propagate in some of the large dimensions. Let us
collectively refer to the light bulk fields as $\phi$. Since the
branes act as coherent sources for $\phi$, they set up a $\phi$
profile in the bulk which is the same as that of a point source
for $\phi$ in the $t$ directions transverse to the source branes
that $\phi$ propagates in.  The precise nature of the dependence
of brane quantities on the volume of the bulk then depends on the
$\phi$ propagators in $t$ dimensions: for $t>3$, the Green's
functions fall of as $r^{2 - t}$ so there is only power suppressed
sensitivity to $V_{bulk}$. For $t=2$ the propagator is a logarithm
and this is precisely what we are looking for. Finally, for $t=1$
there can be power IR divergences (which can have interesting
physical consequences see e.g. \cite{1d}), but which are not of
interest to us here.

We are thus lead to consider theories with light fields which can
effectively propagate in two transverse dimensions. For
simplicity, in the rest of this paper we will assume that there
are only two large dimensions, although we emphasize that
this is not a necessity.  Our picture is summarized in Fig.~1.

\begin{figure}
\PSbox{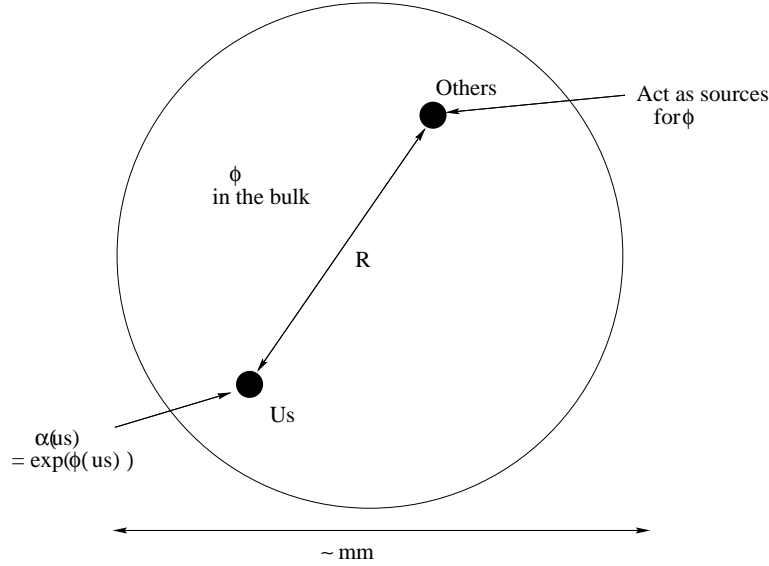 hscale=48 vscale=48 hoffset=80 voffset=-40}{8cm}{5cm}
\vskip.4in
\caption{\it The picture for `running' from the infrared. 
The gauge coupling on our brane is determined by the value of a bulk field
$\phi$ evaluated at the position of our brane. Other branes in the bulk 
a distance $R >> M_s^{-1}$ away act as sources for $\phi$, 
and if the transverse co-dimension relative to the source brane is 2,
then the value of $\phi$ on our brane 
can be logarithmically sensitive to $R$. 
In principle, this logarithmic profile of $\phi$ can mimic field theoretic 
`running' to an energy scale $R M_s^2$ far above the string scale.}
\end{figure}

The
gauge coupling on our brane (evaluated at the cutoff $\sim M_s$)
is the vacuum expectation value of some bulk field smeared out
over a region of size $\sim l_s= M_s^{-1}$ around our brane (as a
consequence of the non-trivial form-factor of the brane):
\begin{equation}
\alpha^{-1} (M_s) = e^{-\phi(us)} .
\end{equation}
However, the presence of other branes far away in the bulk can
modify the value of $\phi$ evaluated on our brane as compared to
the value $\phi_0$ asymptotically far away;
\begin{equation}
e^{-\phi(us)} = e^{-\phi_0} - \frac{c}{2 \pi} \mbox{log} (R M_s).
\end{equation}
For two extra dimensions, $R^2 M_s \sim M_{pl}$, so
\begin{equation}
\mbox{log} (R M_s) = \mbox{log}\left(\frac{M_{pl}}{M_s}\right),
\label{IR}
\end{equation}
and therefore the correct magnitude logarithm between the weak and
gravitational scales is indeed present in the theory. In fact, for
two large extra dimensions, the phenomenological constraints are
tight enough to force $M_s$ at least up to $\sim 50$ TeV. Then,
the logarithm $\log(M_{pl}/M_s)$ is closer to the desired
$\log(M_{GUT}/$TeV), although accurate predictions must of course
also then take into account the usual QFT running between $M_Z$
and $M_s$. Another natural possibility is that the distant branes
are not maximally removed from us in the extra dimensions, so that
$R$ is somewhat smaller.

Note that we are not restricted to a total of $n=2$ large dimensions where
gravity propagates. For instance, suppose we have $n=4$ large dimensions, 
and we are located at the intersection of two orthogonal 5-branes. Fields
that live on the 5-brane propagate in effectively 2 large dimensions and
can vary logarithmically. If the profiles on the two 5-branes are identical 
we can expect to obtain a 
logarithm $ 2 \times$ log$(RM_s) \sim$log$(M_{Pl}/M_s)$, where we have used 
$M_{Pl}^2 \sim R^4 M_s^6$. Clearly this idea generalizes to $n=6$ extra
dimensions as well.  

The fact of logarithmic variation is not enough however. In order
to reproduce the correct `running' of the gauge coupling, the
coefficient $c$ in Eqn.(\ref{IR}) must equal the $\beta$ function
coefficient of the gauge theory localized on our brane. At first
sight, such an equality seems unlikely. After all, $c$ is
determined by the way in which the {\it distant} branes couple to
the light bulk fields; why should this be related to the $\beta$
function of the gauge group localized on {\it our} wall? This
seems to require a miracle.

\section{IR `running' from $N=2$ $D$-brane constructions}

Precisely such a `miracle' is, however, well known to occur in
stringy brane constructions of gauge theories with 8 supercharges.
A simple example is provided by the gauge theories living on D3
branes probing the geometry of parallel D7/O7 configurations \cite{DL}.
(See Fig.~2)
Let the D7/O7 planes fill out the 1,..,7 directions and the D3
brane fill out 1,..,3, and define $w = x^8 + i x^9$ labeling the
space transverse to the 7 branes.  Specifically, put an O7$^-$
plane at $w=0$, and place 4 D7 branes at $w_i$ ($i=1,..,4$). With
4 D7's, the total RR charge and tension of the D7/O7 branes
cancel, so that for $|w| \to \infty$, all the bulk fields approach
constant asymptotic values; in particular the complex coupling
$\tau = a + i e^{-\phi}$ of the type IIB string approaches $i e^{-
\phi(\infty)}$ at infinity.  Now, put a single D3 brane at $w$.
This configuration leaves 8 supercharges invariant, and the
resulting gauge theory living on the D3 brane is $N=2$ in four
dimensions. When all the branes sit on top of each other, the
resulting gauge theory living on the $D3$ brane is an $Sp(1) =
SU(2)$, $N=2$ gauge theory, with the vector multiplet coming from
the 3-3 strings and 4 massless hypermultiplets from the 3-7
strings. This theory is conformal; the one loop beta function coefficient 
is $b=0$. This statement has a
counterpart from the long-distance gravity point of view: with
$w_i$ = 0 all the RR charges and tensions of the branes cancel
against the orientifold, so that there is no variation of the bulk
fields in transverse space. Now suppose we move $f$ of the D7
branes very far away, i.e. $|w_i| = R \gg l_s$. The resulting
gauge theory now has $(4 - f)$ hypermultiplets and beta function 
$b = f$. Now the tensions and RR charges no longer cancel locally, and there
will some variation of the light bulk fields, although of course
asymptotically for $|w|\gg R$ the fields approach their fixed
values at infinity. It is straightforward to calculate the profile
for $\tau$ set up by this configuration of 7 branes. We are in
fact only interested in the long-distance behavior away from the
branes (i.e. only the logarithms), so the full $F$ theory
computation~\cite{Ftheory}
is not needed for our purposes. The result is 
find\footnote{From the full $F$-theory result the non-perturbative
correction to this result can easily be extracted.  They are
negligibly small for a theory that has fundamental coupling of
order the usual GUT coupling.}
\begin{equation}
\tau(w) = \tau(\infty) + \frac{i}{2 \pi}  \left( \sum_{i=1}^{4}
\log (w - w_i) - 4 \log(w)\right).
\label{tauresult}
\end{equation}
This in fact directly follows from the fact that the
seven branes act as vortex solutions for the axion $a=\Re(\tau)$.
This result precisely matches what we expect for the field theory
running of the gauge theory with hypermultiplets of mass $w_i M_s^2$.
The amazing thing of course is that this result of the
long-distance gravity theory (which is valid for $|w_i| \gg
l_s$) reproduces the naive extrapolation of the field theory
running, despite the fact that the field theory description is
valid only for $|w_i|\ll l_s$.

\begin{figure}
\PSbox{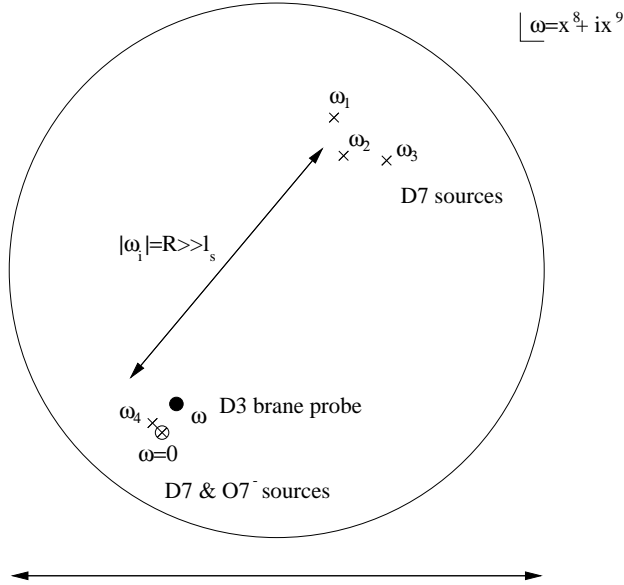 hscale=40 vscale=40 hoffset=80  voffset=-80}{8cm}{5cm}
\vskip .9in
\caption{\it An $N=2$ SUSY example where the bulk SUGRA 
equations reproduce the (3+1)-d QFT holomorphic gauge coupling running
on a probe D3 brane located at $w$.  Source D7 branes are located
at positions $w_i$ in the $w=x^8 +i x^9$ plane, and an O7$^-$
orientifold plane is
located at $w =0$.  At long distances $\gg R$ the total RR charge and
tension of these D7/O7 branes cancel and there is no variation in the
bulk fields.  In this figure we have moved $f=3$ of the D7 branes
far away from the D3 and O7 brane, and so from the D3-brane
gauge theory perspective 3 $N=2$ hypermultiplets gain a large mass 
$m=R M_s^2$.}
\end{figure}

Why does this happen? How can a classical supergravity calculation
reproduce the (quantum) running in a gauge theory? That classical
gravity effects are equivalent to quantum gauge theory effects is
a consequence of closed/open string duality in string theory: The
world-sheet that represents a tree-level closed string exchange
between the well separated branes can also be interpreted, via a
re-labeling of $\xi_1$ and $\xi_2$ on the string worldsheet, as a
1-loop open string diagram of states on the branes that correspond
to strings stretched between the branes. This is illustrated in Fig.~3.
However, this alone is
not enough to explain our miracle. At long distances, the
tree-level closed string diagram is indeed well approximated by
supergravity, but closed/open duality only tells us that this is
equal to the {\it full} one-loop open string calculation, which sums
both the lowest string excitations (which are the field-theory
degrees of freedom) as well as all the massive open string
excitations whose mass depends on $M_s$. In order for the field
theory and gravity calculations to agree, it must be that sum over
massive string excitations gives no corrections to $\tau$. In the
case of theories where the lightest stretched string states
preserve 8 supercharges, this is in fact guaranteed by BPS
considerations~\cite{DL,earlier}.
The lightest states are BPS states with mass and
(NS-NS) charge both linearly increasing with the string length
($M=Q$). The supersymmetry multiplets that these BPS states
fill-out are $N=2$ representations, while the excited strings with
oscillator contributions to their mass are not BPS ($M>Q$) and
group themselves into $N=4$ multiplets, and therefore do not
correct $\tau$.

\begin{figure}
\PSbox{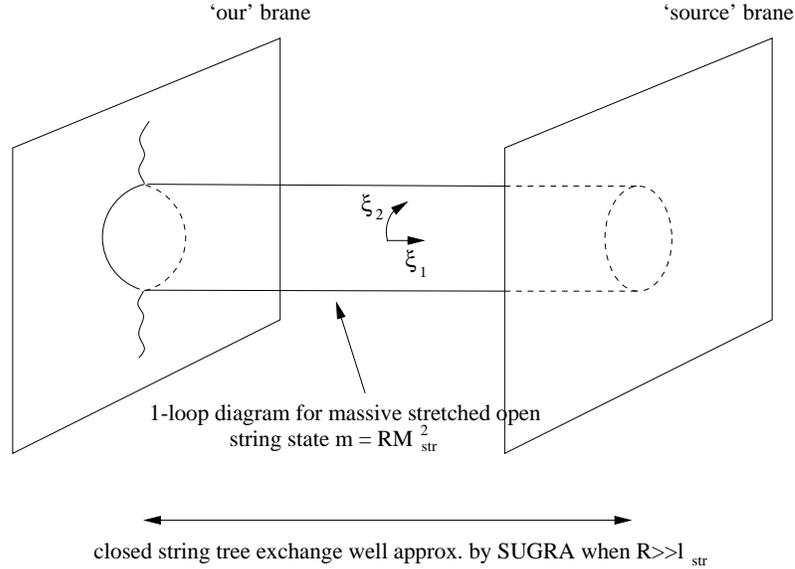 hscale=40 vscale=40 hoffset=80 voffset=20}{8cm}{5cm}
\smallskip
\caption{\it The general picture describing the duality between
1-loop open string and tree-level closed string diagrams.  The worldsheet
coordinates for the string
are $\xi_1$ and $\xi_2$.  If $\xi_1$ is taken to be the worldsheet `time'
coordinate then this diagram represents the exchange of a closed string state
between our brane and a `source' brane in the bulk.  If the separation
$R$ of the branes is large $R\ell_{\rm s}\gg 1$, then this amplitude is well
approximated by the (classical) bulk supergravity solution in the presence
of the brane source.  On the other hand, if $\xi_2$ is taken to be the 
worldsheet `time' coordinate then this diagram represents the 1-loop
contribution of a massive stretched open-string to the
2-point function of our brane-localized theory.  In special cases
(described in the text) only the zero-mode stretched string state
contributes, the higher string
oscillator states canceling, and the usual QFT beta-function
is reproduced.}
\end{figure}

Notice that it is crucial here that the field theory is
superconformal at the origin of moduli space $|w_i| = 0$, and that
moving the branes away to non-zero $|w_i|$ is a soft breaking of
the superconformal symmetry: The heavy hypermultiplets of mass
$|w_i| M_s^2$ act as regulators for the low energy field theory,
and the field theory is finite by itself.  Ordinarily, low-energy
field theory computations have logarithmic divergences which are
cutoff by the softness of string theory at the {\it string} scale.
If we are able to {\it ignore} string oscillators, it had better
be that the field theory yields finite answers by itself.

The expression Eqn.(\ref{tauresult}) is valid when the two transverse
dimensions, $x^{8,9}$ to the D7/O7 planes are non-compact. 
For realistic models we are interested in {\it compactifying} the two
transverse dimensions with a large size $R$ in order to reproduce
the usual $1/r^2$ law for gravity at long distances, so let us
consider what happens with this model when the two transverse
dimensions to the 7 branes are compactified on a large torus with
equal radii $R$.  First note there
is no obstruction to compactification since
the total RR charge and tensions of the 7 branes cancel. For
simplicity, we put the D3 brane at the origin, the O7 at
$w=w_O$ and the D7's at $w=w_{Di}$. We can enforce periodic
boundary conditions by including (same sign) image charges for the
O7 (D7$_i$)'s located at $(m + in)R + w_{O(Di)}$ with $m,n$
integers. It is then important to check that the effect of these
image charges is a small perturbation on top of what we have seen
in the non-compact limit. This must be the case as long as the
$O7/D7's$ are sufficiently far from the ``edges'' of the torus,
i.e. as long as $|w_O|/R, |w_{Di}|/R \ll 1$. Indeed, the exact
expression for $\tau$ evaluated at the position of the $D3$ brane
is
\begin{equation}
\tau(0) = \tau(\infty) + \frac{i}{2 \pi}
\sum_{m,n=-\infty}^{\infty} \sum_{i=1}^{4} \mbox{log} \left(
\frac{m + in + \frac{w_{Di}}{R}} {m + in + \frac{w_O}{R}} \right) .
\end{equation}

The term with $m=n=0$ is just the piece we already have in the
non-compact limit, and it is trivial to Taylor expand the
remaining contribution to find the shift
\begin{equation}
\delta \tau (0) = c \sum_i \left[\left(\frac{|w_{Di}|}{R}\right)^4 -
\left(\frac{w_{O}}{R}\right)^4
\right]
\end{equation}
where $c$ is an $O(1)$ numerical factor. 
We expect this correction to hold in {\it any} realization of our logarithmic
``running'' scenario, since it takes into account the correct Green function 
on the compact space. Already, for $(|w|/R)
\sim 1/3$ this correction is beneath the percent level and will be
irrelevant in any potentially realistic model given the accuracy
to which the SM couplings have been measured.

It is also instructive to consider what happens when only one of the
transverse dimensions is compactified on a circle of radius $r$.
At distances $|w|\gg r$, there is effectively only one transverse
dimension, and $\tau$ varies linearly with $|w|$ rather than
logarithmically. This is simply reflected in the field theory: The
3-7 strings can wind $n$ times around the compact dimension,
giving a tower of (BPS) winding states with masses $n/r$ which are
fundamental hypermultiplets under the $SU(2)$ gauge group.
Including the effect of this entire tower of states changes
logarithmic to power-law running in the field theory. Of course in
field theory we must include in the running not only these states,
but ${\it all}$ states lighter than the `UV cutoff' $|w| M_s^2$.
However, as before the excited string oscillator states are non-BPS
and form themselves in effective $N=4$ SUSY multiplets.  
In the cases where the string excitations make no
contribution to coupling renormalization, we can do
a field theory calculation, keeping all the field theory states lighter than 
the field theory UV cutoff, despite the fact that this cutoff 
is far above the string scale.

This example illustrates that it is indeed possible that field
theory ``running" into the UV is exactly reproduced by logarithmic
variation of bulk fields in the IR.  However, there are two
important issues to be addressed in moving towards more realistic
theories:
\begin{itemize}
\item
How crucial is $N=2$ SUSY? Are there $N=1$
models with the same property?
\item
It is not enough to reproduce the correct `running' for
the couplings; what is the rationale for the gauge couplings to
appear to {\it unify} at $M_{pl}$?
\end{itemize}

This latter question is quite serious and we will address it
first. It is easy to see that the usual argument for unification
in string theory must be lost in the picture we are discussing.
The standard reason for unification is that there is a single
field, the dilaton, whose vacuum expectation value sets the gauge
couplings for all gauge group factors, enforcing equal gauge
couplings at the string scale.  In the present case with a low
string scale, if the gauge couplings of the SM gauge groups living
on our wall are set by a single bulk field $\phi$, then no matter
what other sources $\phi$ has, the gauge couplings are still
guaranteed to unify at the string scale, which is now $\sim$ TeV!
Moreover, we can't have `our' branes, on which the SM gauge group
is realized, significantly separated from each other in the bulk
since this would lead to e.g., very massive $(3,2)$ states under
$SU(3)\times SU(2)$.  Thus we cannot realize different values
of the low-energy SM gauge couplings from evaluating a single 
bulk field at different points in the bulk.
An obvious way around this is if the different gauge couplings
$\alpha_i$ are given by the vacuum expectation values of different
bulk fields $\phi_i$. But then why should we have unification?
Clearly a new rationale is needed.

\section{Unification from Symmetries in the Bulk}

We now give an example of $N=2$ theories where the gauge couplings
unify and the rationale for unification is a geometric symmetry of
the brane configuration in the far IR. These are based on the
Hanany-Witten~\cite{HanW} construction of $N=2$ theories which we
briefly review here. In the simplest set up there are two NS5
branes, filling out 12345 and localized at $x^{7,8,9}=0$,
with one located at
$x^6= 0$ and the other at $x^6 = l$. Suspended between them are
$N_c$ D4 branes located at $x^{4,5,7,8,9}=0$ filling
out $123$ and spanning $x^6=0 \rightarrow
l$. In the absence of the NS5 branes, the gauge theory living on
the D4 branes is a (4+1)-d gauge theory with 16 supercharges.
The boundary conditions of the D4 branes ending on the NS5 branes
reduces the SUSY down to 8 supercharges, and at long distances
compared to $l$ the theory is a (3+1)-d gauge theory with 8
supercharges, which is an $N=2$ $U(N_c)$ gauge theory. The (3+1)-d
gauge coupling is given by reduction from (4+1) to (3+1)-d
\begin{equation}
\alpha_{(3+1)}^{-1} = l \alpha_{(4+1)}^{-1}  =
l M_s {\alpha_s}^{-1}. \label{gau}
\end{equation}

Moving the positions of the D4's corresponds to moving along the
Coulomb branch (adjoint Higgsing) of the $U(N_c)$ gauge theory.
Further suspending semi-infinite $N_f$ D4 branes off the NS5
branes, located at $x^4 + i x^5 = v_i, i=1,..,N_f$ adds $N_f$
hypermultiplets of mass $M_s^2 |v_i|$, coming from the 4-4 strings
stretching between the semi-infinite and finite length D4 branes.

The $D4$ branes ending on the NS5 branes are under tension and
therefore bend the NS5 brane in the $x^6$ direction. The
end-points of the D4 branes are three dimensional, and so are of
co-dimension 2 inside the NS5's.  Therefore we expect that this
bending will be logarithmic. It is well known \cite{MQCD} that
this bending encodes the $\beta$ function of the $N=2$ gauge theory.
To see this physically, imagine first attaching $N_c$
semi-infinite D4 branes on either side of the NS5 branes $x^{4,5}=
0$ (similar to the picture of Fig.~4).
Clearly, the force on each of the NS5 branes
cancel from the left and the right, and so there is no bending of
the NS5 branes: Let us denote the distance between the NS5 branes
as $l_0$. Reflecting the non-bending NS5's, we have an $N=2$
$U(N_c)$ gauge theory with $2 N_c$ hypermultiplets, which is
conformal. Now consider moving some of the semi-infinite D4's away
from the origin; for simplicity move all $N_c$ of the ones on the
left and $N_c - N_f$ of the ones on the right a distance $R$ away.
The resulting gauge theory near $x^{4,5}=0$ is $N=2$ $U(N_c)$ with
$N_f$ hypermultiplets, which has one-loop beta function $b_0 =
(2N_c - N_f)$. Now that the forces on the NS5's no longer cancel
locally, they will bend. However, at distances $\gg R$, the net
bending will cancel and they will still asymptote to being
parallel and a distance $l_0$ apart. We can then sensibly ask for
the distance $l$ between the NS5 branes, at $x^{4,5}=0$, as a
function of the asymptotic distance $l_0$ and $R$; the answer is
\begin{equation}
l = l_0 - \frac{(2N_c - N_f) \alpha_s}{2 \pi} \mbox{log}\left(R/l_s\right).
\label{l}
\end{equation}
It is easy to understand the sign and dependence on $N_c, N_f$ in
the above: the D4's pull on the NS5's, so they are closer to each
other at $x^{4,5}=0$; the D4's between the NS5's pull each of them
inward by an amount proportional to $N_c$ while the D4's on the
outside pull outward by $N_f$, so that the net pull is
proportional to $(2 N_c - N_f)$.

Using the relation Eqn.(\ref{gau}), we find that the value of the
gauge coupling of the gauge theory at $x^{4,5}=0$ varies with $R$
as
\begin{equation}
\alpha^{-1}_{(3+1)} = \alpha_s^{-1}  - \frac{(2N_c - N_f)}{2 \pi}
\mbox{log} \left(R/l_s\right).
\label{run}
\end{equation}

This is precisely the `running' expected only keeping field
theory states. In QFT language, at energies far above $R M_s^2$
the theory is conformal and the coupling doesn't run, its value
being given by $\alpha_s$. Beneath the mass of the $(2N_c - N_f)$
hypermultiplets, the theory runs exactly according to
Eqn.(\ref{run}). Once again, the $(2 N_c - N_f)$ D4 branes that
have been moved away act as `regulators' of the theory left
behind. From the long-distance, gravity point of view they serve
as `IR regulators', canceling the variation of bulk fields so
they asymptote to well-defined values in the deep IR. From the
gauge theory point of view, the low-energy theory has been
embedded inside a softly broken superconformal theory; the massive
hypermultiplets coming from the distant D4's regulate the
low-energy theory, and the superconformality of the full theory
ensure a well-defined value of the gauge coupling in the deep UV.

We can now present our toy example of unification. We will
compactify the $x^6$ direction on a circle \footnote{For simplicity, we
will not compactify any of the other spatial dimensions. Even though 
we do not recover 4D gravity at long distances, doing this allows us to 
most clearly illustrate the picture for gauge coupling unification far 
above the string scale.}, and place 3 NS5's
equally spaced on the circle as in Fig.~4. 
\begin{figure}
\PSbox{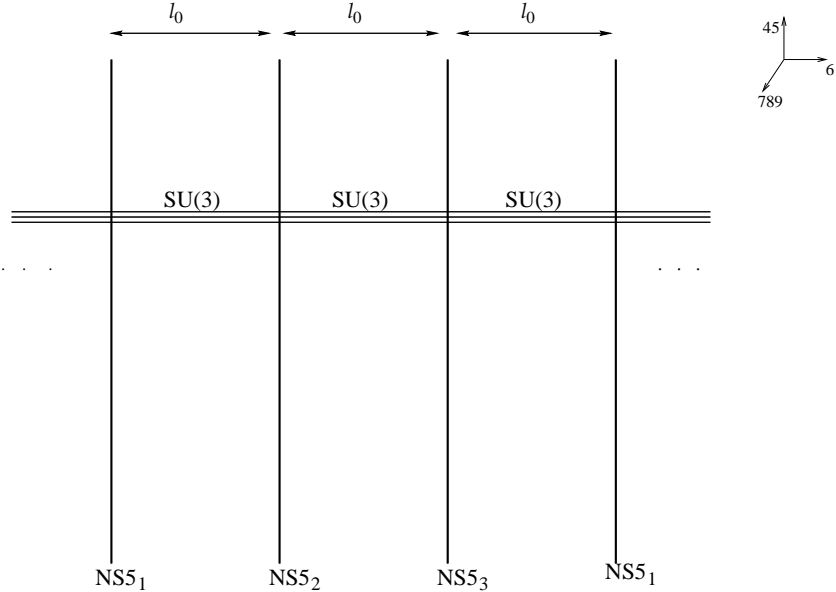 hscale=44 vscale=44 hoffset=50 voffset=-45}{8cm}{5cm}
\vskip .5in
\caption{\it A toy $N=2$ theory with unification far above
the string scale based on the Hanany-Witten set-up.   
The thick lines are NS5 branes and the thin ones are D4 branes. 
The $6$ direction is compactified on a circle, 
which we indicate by periodically
repeating the configuration. The gauge group 
is $SU(3)^3$ with hypermultiplets transforming as 
$(3,\bar{3},1) + (1,3,\bar{3}) + (\bar{3},1,3)$. The NS5's are equally 
spaced so the three gauge couplings are identical.
The forces on the NS5's due to the D4's
cancel locally so there is no bending of the 
NS5's.} 
\end{figure}
In the
supersymmetric limit, there are no forces between the NS5's and we
could place them with any relative spacing we please, but for the
moment let us place them in the most symmetrical arrangement.
Between each of the NS5's suspend 3 D4 branes. The theory is then
$N=2$ $SU(3)^3$ with hypermultiplets in the $(3,\bar{3},1) +
(1,3,\bar{3}) + (\bar{3},1,3)$ representation. This particle
content is an $N=2$ version of `trinification'.
In the limit where
all the D4's sit at $x^{4,5}=0$, all the forces on the NS5 cancel
and they do not bend. Again, reflecting this, the $N=2$ theory is
conformal. Let us now take one of the D4's between NS5$_{2,3}$ and two
between NS5$_{3,1}$ a distance $R$ away as in Fig.~5. 
\begin{figure}
\PSbox{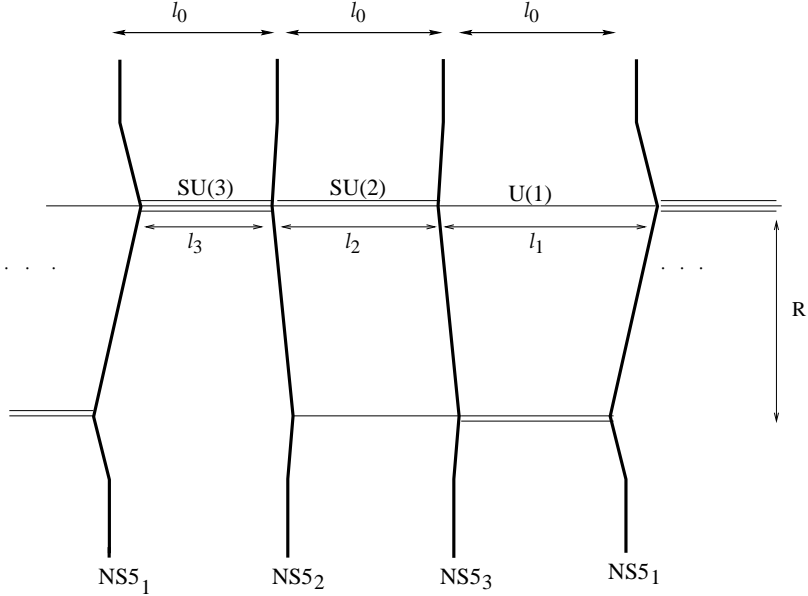 hscale=44 vscale=44 hoffset=50 voffset=-45}{8cm}{5cm}
\vskip .5in
\caption{\it Moving some of the 
D4's a distance $R$ away leaves an 
$SU(3) \times SU(2) \times U(1)$
 gauge group living on the remaining branes.
The forces on the NS5's no longer cancel locally and they bend. We see
that $l_3 < l_2 < l_1$, so $g_3 > g_2 > g_1$.}
\end{figure}
The
resulting gauge theory at the origin has gauge group $SU(3) \times
SU(2) \times U(1)$, with hypermultiplets in the $(3,2,0) +
(\bar{3},1,*) + (1,2,*)$ representation. This is qualitatively
similar to a one generation MSSM, except the hypercharges are
wrong and the theory is N=2. Nevertheless, we can ask about gauge
coupling unification in this toy world. The forces on the NS5's no
longer cancel locally so they will bend locally as in Fig.~5.
However, at distances much larger than $R$, the splitting between the 
D4's can not be resolved and the NS5's flatten out and continue to be 
equally spaced. We know that the bending precisely reproduces the field theory
running in these models. The only question is what the deep UV
value is for the gauge coupling. Because we have arranged for the
NS5's to be equally spaced in the deep IR, in the field theory
this corresponds to the boundary condition that the {\it couplings
unify at the large mass scale $R M_s^2$}. The rationale for
unification at a scale far above the string scale is a geometrical
$Z_3$ symmetry of the brane configuration at large distances. While the usual
picture for unification invokes enhanced symmetries at {\it short} distances, 
here the rationale for unification is an enhanced geometrical symmetry of the 
brane construction viewed from {\it large} distances in the bulk (Fig.~6).
\begin{figure}
\PSbox{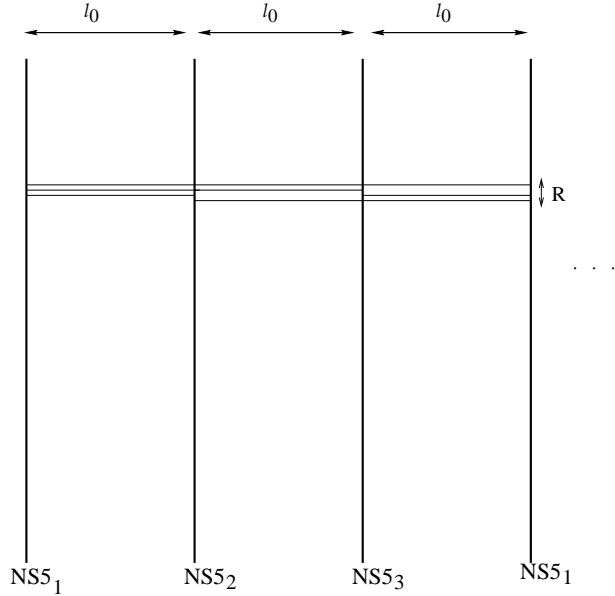 hscale=44 vscale=44 hoffset=50 voffset=-45}{8cm}{5cm}
\vskip .5in
\caption {\it The configuration of Fig.~5 viewed from afar. 
The NS5's flatten out and the $Z_3$ geometrical symmetry of the original 
brane configuration is regained at large distances in the bulk. This provides
a rationale for the apparent unification of the gauge couplings at
a scale much higher than the string scale}
\end{figure}

Notice that this model realizes the general possibility we
mentioned earlier for evading the equality of gauge couplings at
the string scale. At distances larger than the radius of the
$S^1$, we only see that we have the two large dimensions
$x^{4,5}$. The three gauge couplings of $SU(3) \times SU(2) \times
U(1)$ are then indeed given by the values of three bulk fields
corresponding to the inter-NS5 brane distances, and this
interpretation gives us the geometric rationale for unification.

But why should the NS5 branes be placed symmetrically around the
circle? We can 
not fully answer this question until we address SUSY breaking;
nevertheless, we
can at least state some reasonable dynamical assumptions that
generate the desired outcome. If SUSY is broken only on some
branes, then the bulk will still be highly supersymmetric, and we
can expect the brane bending to remain unchanged. On the other
hand, there will now be forces between the NS5 branes, and the
$Z_3$ symmetric configuration will at least be an extremum of the
energy; we need this extremum to be a local minimum. 

It is amusing to think about experimental signatures of this toy
model.  Since the gauge theory becomes effectively (4+1)-d at
energies above $l^{-1}$,there will be KK excitations at this
scale.  But then why do we continue to get log (as opposed to power
law) running?  The reason is that the 16 supercharges of the theory
on the D4's was broken to $8$ supercharges by the boundary
conditions imposed by their ending on the NS5's.  In (3+1)-d
language, we started with an $N=4$ particle content which in $N=2$
language consists of a vector multiplet $V$ and a hypermultiplet
$\Phi$ both in the adjoint representation.  The boundary conditions are
free for the $V$ and fixed for $\Phi$ at $x^6=0,l$. This projects
out the zero mode of $\Phi$ but leaves that of $V$, so the
massless spectrum is $N=2$.  But the massive modes group themselves
into $N=4$ multiplets, and therefore make no contribution to the
$\beta$ function.  There are three different 
KK towers for each of the $SU(3)$, $SU(2)$ and $U(1)$ factors, 
corresponding to the three
strips between the NS5's.  The KK masses then come in the ratio
\begin{equation}
M_{3} : M_{2} : M_{1} = \alpha_3 : \alpha_2 : \alpha_1,
\end{equation}
so this would be a ``smoking-gun" signature of this framework for
unification in this world.

\section{$N=1$ models}

We now turn to the issue of whether $N=1$ models can be
constructed where the field-theory running is reproduced by
supergravity. We do not have a general set of rules for when this
is guaranteed to happen, but will present a class of models which
work. First of all, a good place to look are brane constructions
of $N=1$ theories which are superconformal at the origin of moduli
space. The superconformality guarantees that the field theory
calculation of the running can be finite all by itself without
requiring string theory to cut it off, so that there is at least a
hope that string oscillators can be ignored in the tree-level
close string/1-loop open string comparison. Perhaps the simplest
possibility is to consider orbifolds of $N=2$ models. Indeed,
orbifolds of the Hanany-Witten construction were considered in
\cite{LPT}. Defining $u = x^4 + i x^5$ and $v = x^8 + i x^9$; the
$Z_M$ orbifold considered in \cite{LPT} is $u \rightarrow \alpha
u, v \rightarrow \alpha^{-1} v$ with $\alpha^{M} = 1$. Suspending
$NM$ D4's between the NS5's, the action of the orbifold group
$Z_M$ on the Chan-Paton indices of the $U(NM)$ gauge group is
\begin{equation}
\lambda^a \rightarrow \gamma^{-1} \lambda^a \gamma, \gamma
=\mbox{diag} \left(1_N,\alpha 1_N,\cdots,\alpha^{M-1} 1_N\right).
\end{equation}
The resulting theory is an $N=1$ $SU(N)^M$ theory with chiral
content given by $(N,\bar{N},1,\cdots,1)$ + cyclic permutations.

By lifting this configuration into $M$ theory, Ref.\cite{LPT}
demonstrates explicitly that the bending of the NS5 branes
reproduces the beta function of the $N=1$ theory. We wish to observe
that this can be seen on general grounds and suggests a perhaps
wide class of $N=1$ models with this feature. We find it convenient
again to add `regulator' branes to the system to ensure zero
asymptotic bending in the IR and finiteness in the UV. In the
present case, we add $NM$ semi-infinite D4's on each of the NS5
branes and then orbifold. The resulting $N=1$ theory is an orbifold
of the finite $N=2$ $SU(NM)$ theory with $2NM$ hypermultiplets, and
is easily seen to be a superconformal $N=1$ theory. Now, we move
some of the regulator branes away from $u=0$, which takes them
away from the orbifold fixed point (of course we have to move D4's
and their images together, so they move in groups of $M$). We can
now ask what profile for the bulk fields is set up by these
regulator branes. The crucial point is that since they are away
from the fixed point, they can't act as sources for twisted sector
fields, and therefore they set up exactly the same profile as the
same configuration before the orbifold! Therefore, the
`supergravity' part of the calculation is unmodified. 
On the other hand, as is now well known \cite{orb}, all correlation
functions of an orbifolded field theory (the `daughter' theory) 
agree with the theory before orbifolding (the `parent' theory), 
at the level of planar diagrams, up to a
re-scaling by a factor of $|\Gamma|^{-L}$ where $L$ is the number
of loops, and $|\Gamma|=M$ is the order of the discrete orbifold
group. This result, together with the fact that the `supergravity' 
calculation is unaltered by the orbifold, allows the resulting $N=1$ theory
to inherit the equality between `supergravity' bending and field-theory
running from the $N=2$ theory, as we now show.  (We also note that in the
case where not all branes can be moved away from the fixed point, the 
gauge theory running is still reproduced, at least in the $N=2$ case~\cite{karch},
by the bulk supergravity equations.) 

First of all, the gauge couplings $g^2_p$ and $g^2_d$ of the parent
and daughter theories are related as 
\begin{equation}
\frac{M}{g^2_d} = \frac{1}{g^2_p} .
\label{orbg}
\end{equation}
Note that this implies that the `tHooft couplings of the theories are 
identical. This can most easily be seen by moving the $N$ branes (together 
with all the $M$ images) away from the origin. 
Since all branes are away from the fixed point, 
there is no local way to distinguish the daughter theory from the parent theory
with the same brane configuration. From the daughter theory point of view, 
the $SU(N)^M$ group has been Higgsed to a single $SU(N)$, with gauge coupling
$1/g_1^2 = M/g^2_{d}$. The same gauge configuration in the parent theory 
Higgses $SU(NM)$ to $SU(N)^M$ with gauge coupling $1/g_2^2 = 1/g_p^2$. 
But the gauge coupling of each factor of this $SU(N)^M$
must be exactly the same as that of the $SU(N)$ of the daughter theory, where 
the $M$ factors of the group in the parent theory are interpreted as mirrors 
of the daughter gauge group. Therefore, we must have $1/g^2_1 = 1/g^2_2$ 
and Eqn.(\ref{orbg}) follows. 

Now, we know that the gauge coupling of the parent theory `runs' from the 
IR according to the beta function $b_p$
\begin{equation}
\frac{1}{g^2_p} = \frac{1}{g^2_p(0)} - \frac{b_p}{8 \pi^2} \mbox{log}(R/l_s) .
\end{equation}
Since the bending is unchanged by the orbifold, we also know for the daughter
theory that 
\begin{equation}
\frac{M}{g^2_d} = \frac{M}{g^2_d(0)} - \frac{b_p}{8 \pi^2} \mbox{log}(R/l_s) .
\end{equation}
In order for this to reproduce the field theory `running' for the daughter 
theory it must be that the beta function coefficient of the daughter theory
$b_d$ satisfies $b_p = M b_d$. But this is an immediate consequence of the 
orbifold inheritance results quoted above.  
In particular, the one-loop beta function diagrams are all
planar, so the beta function of each each factor of the daughter 
theory $b_d$ is related to the parent theory $b_p$ as
\begin{equation}
b_d = \frac{b_p}{M}
\end{equation}
This is trivial to see in the
present case: $b_p = 2 N M $ and $b_d = 2N$.

This suggests that the correct `running' can be obtained for a wide class
of $N=1$ theories. One starts with $N=2$ models
where supergravity is guaranteed to reproduce the field theory
``running'', and orbifolds the theory down to $N=1$. If the
regulator branes are moved away from the orbifold fixed point, the
$N=1$ theory can inherit the equality between supergravity and field
theory running from the $N=2$ theory.

Another possibility is that the supergravity does {\it not}
reproduce the $N=1$ running, but that the mismatch comes in
`complete $SU(5)$ multiplets', that is, the mismatch is identical
for $\alpha_{1,2,3}^{-1}$. This would preserve gauge coupling
unification. It is easy to construct a toy example that works in
this way. Going back to the Hanany-Witten set up, we can add $N_f$
parallel D6 branes between the NS5 branes. When the D6's fill out
$123789$, $N=2$ SUSY is still preserved and the 4-6 strings give
$N_f$ hypermultiplets. The bending of the NS5 branes still
reproduces the $\beta$-function of the field theory, the
contribution of the $N_f$ hypermultiplets in the supergravity
description being understood as
follows: The D6 branes are the largest objects in the system and
set up a gravity and dilaton profile in the $456$ space transverse
to them, where the D6's sit at the origin. In this transverse
space, the NS5's fill out a plane $45$, and extremizing the NS5
brane action in the gravity/dilaton profile of the D6 brane causes
logarithmic bending in the $6$ direction, with the correct
coefficient to equal the $-N_f$ contribution of $N_f$
hypermultiplets.

Now consider rotating the D6's so they are parallel to
the NS5 branes, i.e. they fill out $123457$.  It is simple to
check the conditions on the supercharges imposed by this
brane configuration, and discover that only
$N=1$ SUSY is left unbroken.
The particle content is still the same as in
the $N=2$ theory, but the superpotential interaction between the
adjoint in the vector multiplet and the hypermultiplets $\bar{H}
\Phi H$ is switched off, so the interactions only respect $N=1$. In
this configuration, however, the D6 branes {\it do not} bend the
NS5's in the 6 direction; they are parallel to them! The $N_c$
D4's ending on the NS5's still bend them inward by $2N_c$, so the
bending is proportional to $2N_c$ while the $\beta$ function is
still $(2N_c - N_f)$. Therefore the supergravity does not
reproduce the field theory running in this case. However consider
inserting $N_f$ of these D6's between each pair of NS5's in our
trinification model from the previous section.\footnote{Note that
in the $N=2$ configuration the D6's can not be placed between the
NS5's on a compact $S^1$ of a fixed radius, exactly because of the
bending they induce. In our $N=1$ set-up, however, they cause no
bending and can be included.} Now each of the $SU(3)$, $SU(2)$ and
$U(1)$ factors have $N_f$ extra hypermultiplets added, whose
contribution is not reflected in the NS5 bending. But the extra
contribution to the field theory beta function is $N_f$ for all
gauge group factors, so unification is preserved.

\section{Physical couplings and threshold corrections}

So far we have focused on the analogue of 1-loop
``running'' of the gauge coupling coming from the IR. What about
higher-order corrections?

What we have actually been computing is the
logarithmically enhanced contribution to the holomorphic coupling
of the low energy SUSY gauge theory. Of course, the precise way in
which the brane ``thickness'' regulates the value of the gauge
coupling on our brane gives a threshold correction in matching to
the low-energy theory. However, all of these corrections are
due to local physics close to the brane, where the string coupling
is weak, and are not logarithmically enhanced by the size of the
bulk. We do not expect them to be more important than e.g. the
(unknown but small) MSSM threshold corrections. Therefore, our
1-loop expression is an excellent approximation to the holomorphic
gauge couplings of the low-energy theory. Of course, these are not
the physical gauge couplings; the physical couplings are
determined by re-scaling all the fields of the low-energy theory
to go back to canonical normalization \cite{SV}, and are
given by the Shifman-Vainshtein relation
\begin{equation}
\frac{1}{g^2_{ph}} + \frac{2 t_2(A)}{8 \pi^2} \mbox{log}\left(g_{ph}\right) =
\frac{1}{g^2_h} - \sum_i \frac{2 t_2(i)}{8 \pi^2} \mbox{log}\left(Z_i\right) ,
\end{equation}
where $Z_i$ is the wavefunction renormalization of the $i$'th
matter multiplet in the low-energy theory. In field theory, all
the higher loop running of the coupling is contained in the
$\log(g_{ph})$ and $\log(Z_i)$ terms in the above, giving the NSVZ
$\beta$-function.  In our case, we expect that the $Z_i$ are
themselves determined by the vacuum expectation value of a
logarithmically varying bulk field so that
\begin{equation}
Z_i = 1 - \frac{c_i}{8 \pi^2} \mbox{log} \left(R/l_s\right) ,
\end{equation}
with $c_i$ some constants coming from the supergravity solution.
We do not know whether any miracles can guarantee that this
expression reproduces the gauge theory result, although since $Z$
is not holomorphic this seems doubtful. Nevertheless, because only
$\log(Z_i)$ enters in the physical coupling, these corrections
should be of the same order of magnitude as two-loop running in
the field theory, and are also small.

\section{Discussion}

We have seen that the correct `field-theoretic running' of the
couplings may naturally be reproduced by the logarithmic variation
of light fields in the deep IR. However, a new rationale for
unification near the 4-d gravitational scale is required. We
presented a toy model where unification is linked to a geometric
symmetry of the brane configuration in the deep IR. This
correspondence between UV effects in the gauge theory and IR
effects in the gravitational theory gives a fascinating
re-interpretation of what we learn from RGE's. We normally think
that we are limited to doing experiments at low energies, but the
renormalization group allows us to extrapolate the couplings to
much shorter distances, providing an indirect window to this
remote realm. Our toy examples illustrate that sometimes this
interpretation can be misleading. The fundamental short distance
need not be remote at all, it could even be as low as a few TeV.
On the other hand, we are then confined to a 3-brane, and can not
probe {\it large} distances in the transverse space away from our
brane. In our examples, the RGE's, reinterpreted as the result of
logarithmic variation of bulk fields, provide an indirect window
into this new remote realm a millimeter removed in the extra
dimensions. In the old picture, the world look asymmetrical at
large distances but symmetries emerge at short distances. In the
new picture, the world close to our brane looks asymmetrical but
symmetries emerge when we look at large distances in the bulk.

We comment in passing that the ideas presented here can also be
used to provide controllable power-law unification with only one
transverse dimension. The UV sensitivity of usual power-law
unification becomes IR sensitivity in this picture, but since the
IR physics is well-determined in any given model this can be controlled.

Finally, in asymptotically free gauge theories, we are used to
generating energy scales much smaller than the UV cutoff by
dimensional transmutation.  The gauge coupling becomes strong at a
scale $\Lambda$ exponentially smaller than the cutoff and
interesting physics happens. In the standard theories with large
string scale, we expect that some SUSY gauge theory goes strong
and triggers SUSY breaking far beneath the string scale, which
stabilizes an exponentially large hierarchy between the
electroweak and string scale. A similar phenomenon can in
principle occur with low string scale and large extra dimensions.
Suppose that there is a {\it non}-asymptotically free gauge group
living on some collection of branes. Then, the bulk field setting
the gauge coupling becomes strong exponentially far away in the
bulk. This ``dimensional transmutation in the bulk'' naturally
generates an  IR scale exponentially larger than the string scale.
It is tempting to speculate that this scale could determine an
exponentially large effective compactification radius. More generally, 
logarithmic variation of fields in the bulk could force the theory
into strong coupling exponentially far away in the extra dimensions, 
and interesting physics can happen. If an exponentially large radius 
can be generated in this way, we would have a true solution to the hierarchy 
problem, just as compelling as the standard picture in generating the various 
disparate scales observed in nature.

\vspace{0.3cm}

{\bf Acknowledgments}: JMR wishes to thank the Stanford University
Theory Group for their generous hospitality during the main stages
of this work, and both SD and JMR thank the Aspen Center for
Physics for providing a rewarding environment during the
completion of this work.  We would like to thank Ignatios
Antoniadis, Eva Silverstein, and especially 
Martin Schmaltz, Matt Strassler and Angel Uranga for useful discussions.
NAH is partially supported by the DOE under grant DE-AC03-76SF00098 
and the NSF under grant PHY-95-14797, and JMR is supported in part
by the Alfred P. Sloan Foundation.

\def\pl#1#2#3{{\it Phys. Lett. }{\bf B#1~}(19#2)~#3}
\def\zp#1#2#3{{\it Z. Phys. }{\bf C#1~}(19#2)~#3}
\def\prl#1#2#3{{\it Phys. Rev. Lett. }{\bf #1~}(19#2)~#3}
\def\rmp#1#2#3{{\it Rev. Mod. Phys. }{\bf #1~}(19#2)~#3}
\def\prep#1#2#3{{\it Phys. Rep. }{\bf #1~}(19#2)~#3}
\def\pr#1#2#3{{\it Phys. Rev. }{\bf D#1~}(19#2)~#3}
\def\np#1#2#3{{\it Nucl. Phys. }{\bf B#1~}(19#2)~#3}
\def\mpl#1#2#3{{\it Mod. Phys. Lett. }{\bf #1~}(19#2)~#3}
\def\arnps#1#2#3{{\it Annu. Rev. Nucl. Part. Sci. }{\bf #1~}(19#2)~#3}
\def\sjnp#1#2#3{{\it Sov. J. Nucl. Phys. }{\bf #1~}(19#2)~#3}
\def\jetp#1#2#3{{\it JETP Lett. }{\bf #1~}(19#2)~#3}
\def\app#1#2#3{{\it Acta Phys. Polon. }{\bf #1~}(19#2)~#3}
\def\rnc#1#2#3{{\it Riv. Nuovo Cim. }{\bf #1~}(19#2)~#3}
\def\ap#1#2#3{{\it Ann. Phys. }{\bf #1~}(19#2)~#3}
\def\ptp#1#2#3{{\it Prog. Theor. Phys. }{\bf #1~}(19#2)~#3}


\begin{thebibliography}{99}

\bibitem{ADD}
N. Arkani-Hamed, S. Dimopoulos and G. Dvali, {\it Phys. Lett.}
{\bf B429}, (1998) 263.

\bibitem{AADD}
I. Antoniadis, N. Arkani-Hamed, S. Dimopoulos and G. Dvali, {\it
Phys. Lett.} {\bf B436}, (1998) 257.

\bibitem{ADDlong}
N. Arkani-Hamed, S. Dimopoulos and G. Dvali, {\tt hep-ph/9807344}.

\bibitem{AHDMR}
R. Sundrum, {\tt hep-ph/980734};\\
N. Arkani-Hamed, S. Dimopoulos and J. March-Russell, {\tt hep-th/9809124}.

\bibitem{AHD} N. Arkani-Hamed and S. Dimopoulos, {\tt hep-ph/9811353}.

\bibitem{ADDMR}
N. Arkani-Hamed, {\it et al}, {\tt hep-ph/9811448};\\
Other ideas were suggested by K. Dienes, E. Dudas, T. Gherghetta,
{\tt hep-ph/9811428}. 

\bibitem{RS1} 
L. Randall and R. Sundrum, {\tt hep-th/9810155}.

\bibitem{AHS}
N. Arkani-Hamed and M. Schmaltz, {\tt hep-ph/9903417}.

\bibitem{cosmo}
N. Kaloper, {\tt hep-th/9811141};\\
G. Dvali and S.H-H Tye,  {\tt hep-ph/9812483};\\
N. Arkani-Hamed {\it et al}, {\tt hep-ph/9903224}, {\tt hep-ph/9903239}.

\bibitem{DDG}
K. Dienes, E. Dudas and T. Gherghetta, {\it Phys. Lett.} {\bf
B436}, (1998) 55;\np{537}{99}{47};\\
D.~Ghilencea and G.~Ross, {\tt hep-ph/9809217}.

\bibitem{unif}
H. Georgi and S. Glashow, \prl{32}{74}{438};\\
H. Georgi, H. Quinn and S. Weinberg, \prl{33}{74}{451};\\
S. Dimopoulos and H. Georgi, \np{193}{81}{150}; \\
S. Dimopoulos, S. Raby and F. Wilczek, \pr{24}{81}{1681}.

\bibitem{Dbrane}
See for example:\\
J. Polchinski, {\it TASI lectures on D-branes},
{\tt hep-th/9611050};\\ C. Bachas, {\it Lectures on D-branes},
{\tt hep-th/9806199};\\ A. Giveon and D. Kutasov, {\it Brane dynamics
and gauge theory}, {\tt hep-th/9802067}.

\bibitem{HanW}
A. Hanany and E. Witten, \np{492}{97}{152}; {\tt hep-th/9611230}.

\bibitem{MQCD}
E. Witten, \np{500}{97}{3}; \np{507}{97}{658}.

\bibitem{LPT}
J. Lykken, E. Poppitz and S. Trivedi, \pl{416}{98}{286},
{\tt hep-th/9708134}.

\bibitem{DL}
M. Douglas and M. Li, {\tt hep-th/9604041};\\
C. Bachas and C. Fabre, \np{476}{96}{418}.

\bibitem{matt}
A. Hanany, M. Strassler and A. Uranga, {\tt hep-th/9803086}.

\bibitem{orb}
S. Kachru and E. Silverstein, \prl{80}{98}{4855},
{\tt hep-th/9802183};\\ M. Bershadsky, Z. Kakushadze and C. Vafa,
\np{523}{98}{59}, {\tt hep-th/9803076};\\ M. Bershadsky and A. Johansen,
\np{536}{98}{141}, {\tt hep-th/9803249}.

\bibitem{karch}
A. Karch, D. Lust and D. Smith, {\tt hep-th/9803232}.

%\bibitem{fracbrane}
%D. Diaconescu, M. Douglas and J. Gomis, hep-th/9712230;\\ M.
%Douglas and G. Moore, hep-th/9603167.

\bibitem{B}
C. Bachas, {\tt hep-ph/9807415}.

\bibitem{AB}
I. Antoniadis and C. Bachas, {\tt hep-th/9812093}.

\bibitem{earlier}
I. Antoniadis, C. Bachas and E. Dudas,
{\tt hep-th/9906039};\\
L. E. Ibanez, {\tt hep-ph/9905577};\\
E. Halyo, {\tt hep-ph/9905577}.

%\bibitem{radius}
%N. Arkani-Hamed, S. Dimopoulos and J. March-Russell, in
%preparation.

\bibitem{1d}
L. Randall and R. Sundrum,  hep-ph/9905221, hep-th/9906064;\\
M. Goberashvili, hep-ph/9812296.

\bibitem{Ftheory}
E. Bergshoeff, {\it et al.}, \np{470}{96}{113};\\
C. Vafa, \np{469}{96}{403};\\
A. Sen, \np{475}{96}{562}.

%\bibitem{trinification}
%S. Glashow, in {\it Providence Grand Unification} (1984) pp.88. 

\bibitem{antoniadis}
I. Antoniadis, \pl{246}{90}{377};\\
I. Antoniadis and K.~Benakli, \pl{326}{94}{69};\\
I. Antoniadis, K.~Benakli and M.~Quiros, \pl{331}{94}{313}.

\bibitem{SV}
M.A. Shifman and A.I. Vainshtein, \np{359}{91}{571};
N. Arkani-Hamed and H. Murayama, {\tt hep-th/9707133}.

\end{thebibliography}
\end{document}